# Making Web Annotations Persistent over Time

Robert Sanderson
Los Alamos National Laboratory
Los Alamos
NM 87544, USA
+1 (505) 665-5804
rsanderson@lanl.gov

Herbert Van de Sompel
Los Alamos National Laboratory
Los Alamos
NM 87544, USA
+1 (505) 667-1267
herbertv@lanl.gov

## ABSTRACT

As Digital Libraries (DL) become more aligned with the web architecture, their functional components need to be fundamentally rethought in terms of URIs and HTTP. Annotation, a core scholarly activity enabled by many DL solutions, exhibits a clearly unacceptable characteristic when existing models are applied to the web: due to the representations of web resources changing over time, an annotation made about a web resource today may no longer be relevant to the representation that is served from that same resource tomorrow.

We assume the existence of archived versions of resources, and combine the temporal features of the emerging Open Annotation data model with the capability offered by the Memento framework that allows seamless navigation from the URI of a resource to archived versions of that resource, and arrive at a solution that provides guarantees regarding the persistence of web annotations over time. More specifically, we provide theoretical solutions and proof-of-concept experimental evaluations for two problems: reconstructing an existing annotation so that the correct archived version is displayed for all resources involved in the annotation, and retrieving all annotations that involve a given archived version of a web resource.

## Categories and Subject Descriptors
H.5.4 [**Information Interfaces and Presentation**]: Hypertext/Hypermedia – *Architectures, Navigation.*

## 1. INTRODUCTION

Annotation was identified by John Unsworth [25] as a basic function common to scholarship across all disciplines, along with six others such as comparing, discovering and referring. He named these functions "scholarly primitives", being core and atomic elements of the scholarly process. Annotations are used in scholarly communication to organize existing knowledge and to facilitate the creation and sharing of new insights. They can become so important as to have scholarly value in their own right, and hence their transition from paper to digital scholarship has been, and remains today, crucially important.

Most Digital Library (DL) architectures include services that enable users to annotate the managed objects. Existing annotation solutions are often tightly connected both with specific content collections and the DL toolkits that manage those collections. This prevents both sharing of annotations beyond the boundary of collections, and the use of the same tool to annotate collections managed by different digital library solutions.

As DLs move away from being the isolated content silos of the previous decade and towards full web integration, they become more closely aligned with the architecture of the web [14] and Fielding's REST paradigm [9]. As this process matures and the digital objects held by the DL are set free as first class resources on the web, many value-adding components such as those enabling annotation will require rethinking in terms of the architectural foundations of the web.

The web architecture is centered on the Uniform Resource Identifier (URI) that identifies a resource, defined as an item of interest [4]. When a client such as a web browser dereferences the URI of a resource, a representation (in the form of a bitstream plus metadata) of the current state of that resource is returned. As time goes by, a resource keeps the same URI, but its representations are likely to change. To put it simply, when clicking http://www.cnn.com/, the current version of the CNN home page is returned. But, when clicking that same HTTP URI tomorrow, a different version of the home page will be returned.

The problem this architectural design causes for a web-centric annotation framework should be clear. An annotation made about a paragraph of today's CNN homepage will naturally be expressed in terms of the homepage's URI. However tomorrow, when recalling that annotation, it will be overlaid on a new version of the CNN homepage and will most likely be totally irrelevant. This lack of robustness of annotations over time is clearly unacceptable for web-based scholarship.

The web architecture is of no assistance in solving this issue as it does not consider time at all: once a representation has been returned from a resource, any further operations on that representation are out of scope. Digital Preservation systems, such as web archives, can make each such representation available as a distinct archival resource with its own URI. However, there is no means of automatically discovering the archived resource given only the URI of the resource that was originally annotated. A skilled and knowledgeable human is needed in order to know how and where to search, and even then there is no guarantee that the resource has been archived where the user searches.

Existing annotation frameworks do not make it easy either, as after extensive research none were found with a model that allowed for different component resources to be considered at different points in time. All assumed that the resources would be uniquely identified, and that this would be sufficient. They instead relied on heuristics to re-attach the annotation after the inevitable modifications to the annotated resource had occurred.

This paper explores how a web-centric annotation framework can support the necessary robustness of annotations by making use of the archived representations from the time when the resource was originally annotated. The approach aims to be transparent to end-users and at the same time true to the web architecture's fundamental principles. The investigation is formulated as two parallel research questions:

1. Given an annotation about a web resource, can the appropriate representation to which it pertains be retrieved? For example, can we use the information in an annotation to retrieve an appropriate archived version of last year's CNN homepage to which the content of the annotation is relevant?

2. Given an archived representation of a web resource, can the appropriate annotations be retrieved which apply to it? For example, given that archived version of last year's CNN homepage, can we rediscover the annotations about it?

These questions, more formally defined in Section 4, are considered in the context of two ongoing research projects, the Open Annotation Collaboration [22], and the Memento Project [26] which will be described in detail in Section 3. We describe the proposed solution in Section 4 followed by a practical evaluation and discussion of the challenges involved.

## 2. PREVIOUS WORK

The breadth and depth of research on the topic of annotation is vast and it is beyond the scope of this paper to provide a summary. Instead, we start by highlighting the works of two groundbreaking authors.

More than ten years ago, Cathy Marshall described her study [18,19] of annotations across 410 physical textbooks, and the implications for transferring that activity into the nascent DL paradigm. Marshall considers the form the annotation took and the function of the annotation, identifying reasons for annotating such as "procedural signals", "placemarks", "working problems" and "interpretive activity", and seven orthogonal dimensions of annotation. The work is invaluable for understanding annotators and their motivations, practically providing a checklist for use cases and requirements.

Maristella Agosti has published prolifically on the topic of annotations in the context of digital libraries over the last decade, with more than 15 high quality publications focusing on the topic. Of particular note is the formal model for annotation [1], which exhaustively describes and categorizes the various entities involved in the process of annotation, with reference to Marshall's earlier work. Although the implementation described in [2] is very much that of a service-centric rather than web-centric digital library, the defined concepts are exemplary for ensuring that all of the modeling and architectural requirements are covered.

There have been several web-centric annotation systems, starting with the W3C work on Annotea [15]. Annotea consists of a minimal model described in RDF, plus an extensive protocol for interaction with annotation servers in a REST based fashion. All noteworthy subsequent RDF based annotation systems have been extensions of Annotea. We consider two such systems.

Hunter extends Annotea to allow for multiple annotated resources [23], for example in order to allow the annotation to represent a relationship between two targets. The Vannotea system also allows for multiple media types including video annotation and SVG described media segments.

The second extension is the LEMO framework [12] of Haslhofer and colleagues, integrated into both the FEDORA Digital Library platform and The European Library (TEL). Like Annotea, it has a REST protocol for client/server interaction. In addition it follows the Linked Data [5] guidelines, and provides an extensibility mechanism to allow for complex segment descriptions and multimedia annotations.

None of the previously described work has looked at the challenges of persistence or robustness of annotations and their constituent resources with respect to time caused by following an approach founded on the architecture of the web. Research into robustness has focused primarily on detecting whether the annotation should be relocated within the resource (eg what was paragraph 2 at the time of annotation is now paragraph 3 in the current state), or should be discarded as no longer relevant.

Phelps and Wilensky discussed Robust Locations [20] as implemented in the Multivalent Document system. This allows for re-attachment of the annotation to text or locations within a document using a generalized document model. This research has been extended by Corubolo and colleagues [8] to cross formats and locations, and re-used in subsequent work in the context of DL architectures [3]. Although recent versions of the software have an RDF serialization for the annotation model, it would be difficult to describe the approach as web-centric.

Golovchinsky has also looked at re-attachment of annotations into modified documents [11], including dynamically scaling the marks of annotation such as hand-drawn underlining and highlighting. Brush et al. [6] consider robust annotation positioning at the other end of the complexity spectrum with a very simple sub-phrase discovery based algorithm, and show acceptable results with web pages.

These advances seek to re-attach the annotation to a new version of the annotated document through the use of heuristics, or to discard it if it is no longer relevant. Our web-centric annotation problem is different and hence it requires a different solution. We aim to seamlessly recover the representation of an annotated resource from where it is maintained in a web archive, and use it to redisplay the annotation in its original context.

## 3. COMPONENT SYSTEMS

Before turning to the design for our solution, it is necessary to discuss the two primary components it builds upon. First, in Sections 3.1 through 3.3, the Open Annotation Collaboration (OAC) data model, information architecture and temporal design are discussed, and then Section 3.4 presents the Memento solution for navigation to archived representations of web resources.

### 3.1 Open Annotation Data Model

The OAC model [22] is informed by the previously discussed work, and tries to integrate the various extensions of Annotea into a cohesive whole. Indeed, Jane Hunter and Bernhard Haslhofer have been instrumental in the model design, and Cathy Marshall and Maristella Agosti are members of the OAC advisory board. The web architecture and linked data guidelines are foundational principles, resulting in a specification (in Alpha release at the time of writing) that can be applied to annotate any set of web resources. As such, OAC can be used to annotate web-centric DL resources in the same way as YouTube[1] videos, photos on Flickr[2] or any other resource identified by a URI.

Following its predecessors, the OAC model, shown in Figure 1, has three primary classes of resource. In all cases below, the *oac* namespace prefix expands to *http://www.openannotation.org/ns/*.

1. The *oac:Content* of the annotation (node URI-C). This resource is the comment, metadata or other information that is created about another resource. The Content can be any web resource, of any format, available at any URI. The model allows for exactly one Content, connected using the *oac:hasContent* predicate, with the Annotation as subject and the Content as object.

2. The *oac:Target* of the annotation (node URI-T). This is a resource that the Content is about. Like the Content, it can be any URI identified resource. The model allows for one or more Targets, connected using the *oac:hasTarget* predicate, with the Annotation as subject and each Target as object.

3. The *oac:Annotation* event (node URI-A). This resource stands for the event in which the relationship between the Content and the Target is assigned. That relationship is explicitly stated using a predicate called *oac:predicate*, with the default being *oac:annotates*.

It must be stressed that the Annotation and the Content may be created at different times and by different agents. For example, one might create an Annotation in which someone else's YouTube video annotates a third person's Flickr photo. The Content and Target resources may not even be aware that they are part of the annotation in this case.

When the user annotates a resource, initially the Content may be just a string in their application. Strings are not web resources; they of course do not have a URI. However, there are many services that will happily assign URIs to strings for free, such as Twitter[3], various blog platforms or Google Docs[4]. Henceforth we treat Content nodes as resources on the web with their own URIs, even if they did not start their existence with one.

The Annotation is a conceptual resource that denotes an event in time, and is modeled as a non-information resource [17] that does not have a representation, but instead is described by another resource. In OAC, the Annotation is described by a resource of class *oac:Transcription* (node URI-Trn) which records all of the information about the Annotation and its components, and therefore serves the same purpose for the Annotation as an OAI-ORE [27] ResourceMap does for an Aggregation. A Transcription provides a serialization of the Annotation graph in one of the RDF formats such as RDF/XML. The Annotation is the object of the *oac:transcribes* predicate where the Transcription is the subject.

Many Annotations concern parts, or segments, of resources, rather than the entire resource. While simple segments can be identified and referenced directly using the emerging W3C Media Fragment specification [24], there are many use cases for segments that currently cannot be identified using this proposal. For example, the W3C specification only allows rectangular sections of images to be identified with a URI, but not any arbitrarily defined region. To allow for the identification of any region of interest, the OAC model follows Hunter's [23] extension of Annotea's context node, by defining the oac:*Context* class (node URI-Ctx), which can capture segment and other information about the resources in the context of the annotation. The predicates used for these relationships are *oac:hasTargetContext* and *oac:hasContentContext* from Annotation to Context node for Target and Content respectively, and *oac:contextAbout* from Context to the appropriate Target or Content node.

The description of the segment is captured by a resource of the *oac:SegmentDescription* class (node URI-SD) related to the appropriate Context node via the *oac:hasSegmentDescription* predicate. This description could be an SVG document, or inline SVG content, to describe an area of an image, an XPath or XPointer for XML or HTML, or a more complex description for slices of datasets or other resources.

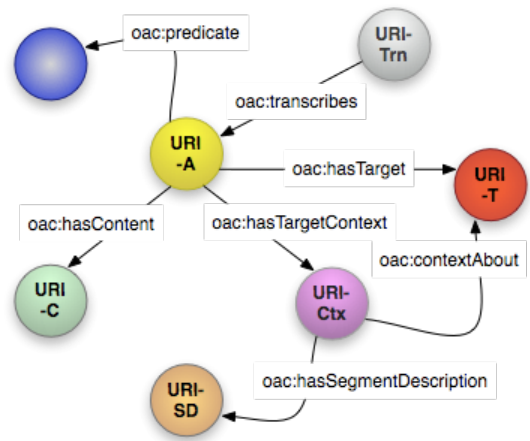

**Figure 1. Basic OAC Alpha Data Model**

## 3.2 Time and Annotations

The way in which the OAC data model handles time is, of course, essential when we consider our research questions of retrieving the correct representation for an annotation, and all annotations that apply to a particular representation.

The Annotation itself is not time dependent; it is an event at a particular point in time and hence it cannot be modified. While it cannot change, the Content and Target resources are very likely to have different states over time and the annotation may or may not apply to these evolving states.

The OAC Data Model distinguishes three different types of Annotation with respect to time: Timeless Annotations, and two Time Dependent Annotation variants. These can be distinguished based on the presence or lack of the *oac:when* relationship, which has a datetime as a value. These types of annotation are depicted in Figure 2.

A Timeless Annotation can be considered as if the Annotation references the semantics of the resources, and is not dependent on the representations of those resources at particular points in time. For example, an annotation that states that "This is the front page of CNN" does not depend on the current state of

---

[1] http://www.youtube.com/

[2] http://www.flickr.com/

[3] http://www.twitter.com/

[4] http://docs.google.com/

`http://www.cnn.com/`, it is about the purpose of the resource identified by that URI. As such, there is no need to situate the resources at a specific moment in time, and therefore there is no use of the *oac:when* predicate. The Content resource in a Timeless Annotation is always applicable to the Target resource(s) and hence the supplied or implicit predicate for the relationship between them always holds true.

Uniform Time Annotations have a single point in time at which all of the resources involved in the annotation should be considered. This is expressed via an *oac:when* property on the Annotation. If this situation is encountered, it signals that the annotation is known to be valid at this point in time but is not necessarily so at other points in time. To continue the CNN example, an annotation about a particular story on the current homepage would need this style of annotation, as the target story will likely not be there within several hours, let alone forever.

Varied Time Annotations also use the *oac:when* property, but it is attached to the Context nodes of individual resources instead of the Annotation. This allows annotations to involve resources that should be interpreted at different times. For example, a blog post written yesterday about the CNN homepage of the previous day would be this type of annotation. Uniform Time Annotations could be re-expressed in this more verbose form, however this requires the construction of a significantly more complex graph, as depicted below.

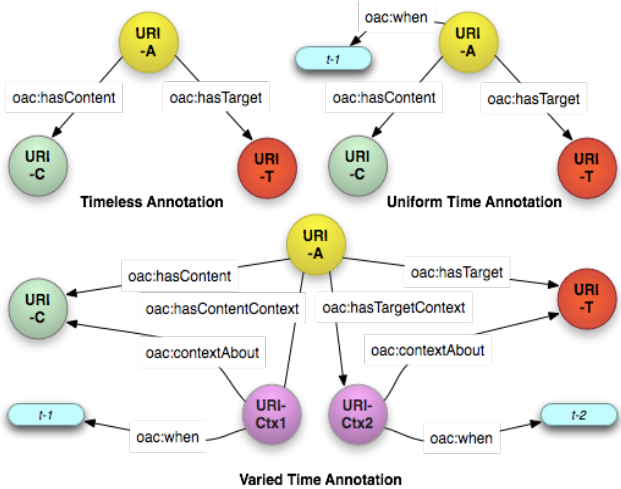

**Figure 2. Timeless and Time Dependent Annotations**

Segment Descriptions can also be web resources in their own right and hence would, in principle, need their own *oac:when* property for the model to be completely customizable with respect to resource state. However, these resources are considered to be immutable as they are very tightly tied to the Annotation. This means that the OAC model assumes that Segment Descriptions are never changed, other than to correct an error.

## 3.3 Open Annotation Architecture

One of the criticisms frequently leveled against Annotea was the tight coupling of the client and server via a protocol that specified how to create, update, retrieve and delete annotations, by using HTTP verbs and RDF/XML as the body of the transactions. This approach has hindered adoption of Annotea, and therefore OAC pursues a publish/discover architecture in which clients and servers are decoupled.

The OAC architecture, contrasted with that of Annotea in Figure 3, attempts to do this by continuing in the web-centric paradigm. The OAC Client makes the Annotation's Transcription available on the web just like any other resource. The Annotation Collection service discovers and retrieves it, then adds it to its database of Annotations. Clients can then interact with the database to search for and retrieve annotations for a particular target resource and optionally filter by time. Additional services such as Aggregators can then be overlaid on this base architecture.

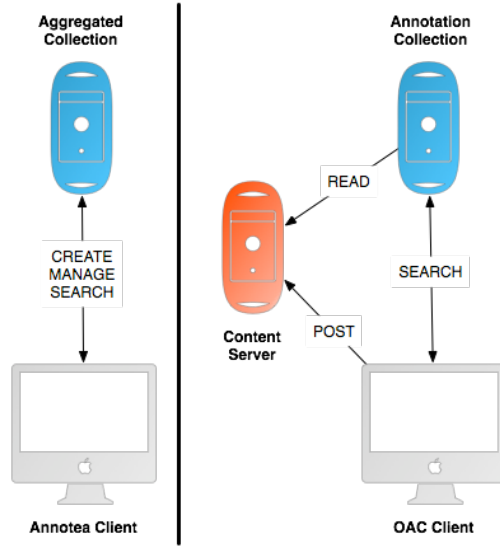

**Figure 3. Annotea vs OAC Architectures**

The primary advantage of this architecture is that all of these interactions use regular web publishing and discovery techniques. The Transcriptions could be written to a blog, and then discovered by the Annotation Collection by subscribing to an RSS/ATOM feed, for example. The client can also interact with the database by using existing techniques such as filling in the blanks in a URL template to get to a list of annotations that involve a specific resource as a Target, as implemented by Google's Sidewiki[5], or via search protocols such as OpenSearch[6] or SRU[7].

Removing the need for authorization hooks in the model is another advantage of the architecture, avoiding the extensive user, group, role and privilege modeling requirements in Agosti's formal approach, or additional systems such as Hunter's use of XACML [16]. If the annotator wants to restrict access to the annotations, then authentication can be added at the Content Server in the same way it is done for any other web resource.

## 3.4 Memento

The second component in the solution is Memento [26]. The key insight is that in order to retrieve an old version of a resource, instead of having to search for it, it would be preferable to simply go to the URI of that resource and request a representation from the past rather than the current representation. Memento achieves this by re-using the existing functionality present in every web browser that allows it to negotiate with the server as to the desired format of the representation to be returned. This functionality is called Content Negotiation [13] and happens all the time on the web, most commonly to present a representation of a resource in a different format to a small form factor client such as a cell phone, as compared to one sent to a regular browser. Memento uses this capability to negotiate about time, rather than media type, and the interactions are depicted in Figure 4.

---

[5] http://sidewiki.google.com/

[6] http://www.opensearch.org/

[7] http://www.loc.gov/standards/sru

Memento introduces a new HTTP request header, currently called `Accept-Datetime`, which contains the desired timestamp for the state of the resource, henceforth called the Original Resource. This header mirrors the existing ones for content negotiation such as `Accept`, `Accept-Language` and so forth. When in Memento mode, instead of retrieving the current representation, a client will go to a TimeGate; a resource that handles the time-based negotiation. The TimeGate could be managed by the same server as the Original Resource, as in the case of a content management system that maintains version information, or run by a third party such as a web archive. Clients can maintain lists of TimeGates to use, and the Original Resource can advertise its suggested TimeGate via a `Link` header entry with a `rel` parameter of "timegate", of the form:

    <URI-TimeGate>;rel="timegate"

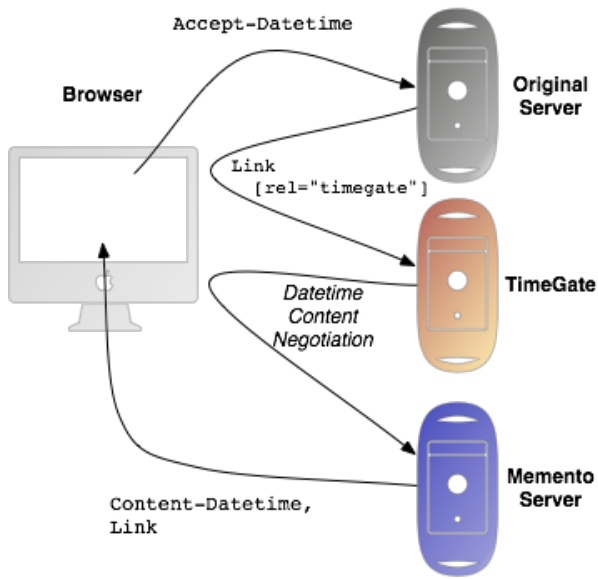

**Figure 4. Memento Architecture**

The TimeGate then processes the timestamp and, following regular content negotiation procedures, attempts to redirect the client to a resource that meets the client's preference for time. If successful, the client arrives at an archival resource (termed a Memento) that has a representation that is the same as the original resource had at the given time. In essence the TimeGate resolves the tuple of (Original Resource URI, timestamp) into the URI for the Memento. If the TimeGate is unable to fulfill the request, it can return an error message explaining the problem, otherwise the client can follow the redirection to retrieve the Memento.

The same requirements for regular media-type content negotiation also apply for datetime content negotiation. The TimeGate must add a `Link` header to its response that contains pointers to the Original URI and to other Mementos and their times. The Memento server must also add a `Content-Datetime` header with a value of the time of the Memento being returned.

In addition, the Memento solution proposes an API for web archives that enables the archive to list all of the Mementos it has for a given Original Resource. This is implemented as a TimeBundle, which is a specialization of an ORE Aggregation, and has serializations called TimeMaps that follow the regular RDF and Atom serialization rules for ORE. To support discovery of these resources, TimeGates and Memento servers should include a pointer to the TimeBundle in the `Link` header.

## 4. CONCEPTUAL SOLUTION

With OAC and Memento, the necessary components are available in order to test the two research questions. The components of OAC Annotations are all web resources that can be archived. Instead of having to search for these by hand, they can seamlessly be discovered by following the Memento HTTP redirections from the Original Resource.

For this solution, we assume that compliant implementations are available which enable retrieval of the Transcriptions using the Target, Content and Annotation URIs as well as their timestamps as search keys. For simplicity, we do not consider Annotations with multiple Targets, as if the solution works for one Target, it will work for more than one. Finally, as Uniform Time Annotations could be re-expressed as Varied Time Annotations and Timeless Annotations are relevant regardless of the specific representation served, we will consider only the Varied class.

Before defining the research questions more formally, we give some definitions:

- A representation obtained from a resource URI-X at time $t_i$ is $rep(URI-X, t_i)$.
- A Memento URI-M for a resource URI-R at time $t_i$ is the resource $URI-M(URI-R, t_i)$, and for any time $t_j$ after $t_i$, $rep(URI-M(URI-R, t_i), t_j)$ must be the same as $rep(URI-R, t_i)$.
- Subscript i, j and k are positive integer values

We now define the research questions more formally as:

1. Given an Annotation URI-A, that involves web resources $URI-R_k$, each with a representation at a point in time $rep(URI-R, t_i)$, can URI-A be reconstructed faithfully at time $t_j$, with $t_j > max(t_i)$?
2. Given a Memento $URI-M(URI-R, t_i)$, can the appropriate Annotations $URI-A_k$ be retrieved that apply specifically to $rep(URI-R, t_i)$, with $t_i <=$ current time?

### 4.1 Mementos for a Given Annotation

The first question to be answered is whether the state of the resources involved in a given annotation can be reconstructed as they were at the time of annotation. For example, given a comment about the CNN homepage at a certain point in time, can we reconstruct both the comment and homepage as they were at that point? The process, described below, is depicted in Figure 5.

A Varied Time Annotation URI-A is created with a Content node URI-C from time $t_2$ which is about a Target node URI-T, from time $t_1$. A Transcription URI-Trn is created that describes URI-A, and stored in a Content Server. URI-C and URI-T are now URI-$R_{1..2}$ from the formal definition. The client alerts the Annotation Collection to retrieve the Transcription from the Content Server.

At any point later in time $t_j$ with $t_j > \max(t_1, t_2)$, the client retrieves an Annotation Transcription from the Annotation Collection. In order to faithfully reconstruct the Annotation, the client must discover the tuples (URI-T, $t_1$) and (URI-C, $t_2$) from the graph in URI-Trn. As the times for the resources are different, they will be recorded using the *oac:when* property on the Context nodes in the RDF graph. These can be extracted by traversing the graph parsed from the Transcription, or via the SPARQL[8] query:

```
SELECT ?uri_r ?ti WHERE {{<URI-A>
oac:hasTargetContext ?ctxt} UNION {<URI-A>
oac:hasContentContext ?ctxt} . ?ctxt a
oac:Context . ?ctxt oac:contextAbout ?uri_r .
?ctxt oac:when ?ti .}
```

For each URI-R and i=1..2, the client sends an HTTP GET request to URI-R with the `Accept-Datetime` header set to $t_i$. The server will redirect each request to a TimeGate URI-TG(URI-R), which in turn performs datetime content negotiation to redirect to a Memento URI-M(URI-R, $t_i$).

Once the tuples have been resolved to the appropriate URI-M, the client retrieves and displays the representation from URI-M instead of the current representation from URI-R.

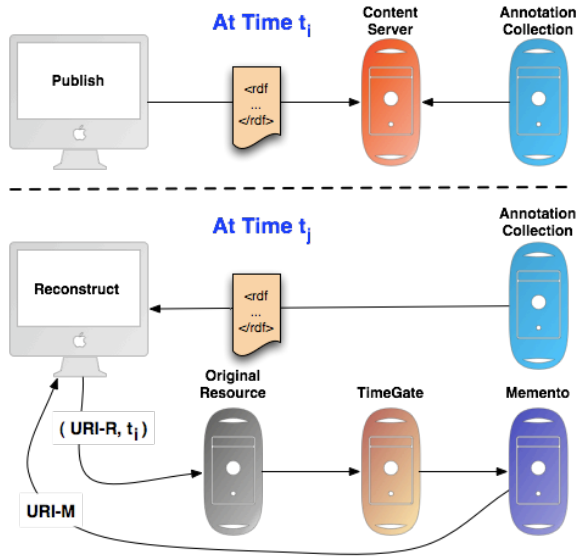

**Figure 5. Memento for Annotation Model**

## 4.2 Annotations for a Given Memento

The inverse of the first question is, given a Memento URI-M or the Original Resource URI-R plus a timestamp, is it possible to discover all of the annotations that apply to that version of the resource? In this case, the user has come to an archived copy of a web page and wants to know what people thought about that version, rather than about the current page served from the Original Resource.

We break the problem down into three steps. The first, and easiest, is given URI-M, is it possible to discover URI-R? Then,

---

[8] http://www.w3.org/TR/rdf-sparql-query/

given URI-R, is it possible to discover the values of i (the time interval) in which rep(URI-M) is functionally equivalent to rep(URI-R, $t_i$)? Finally, we must discover annotations with target URI-R in that time interval from the Annotation Collection.

First we discuss how to discover URI-R and an approximation of the time range, before looking at how to make that more precise.

### 4.2.1 Approximation Approach

To discover URI-R from URI-M, the application can inspect the `Link` header of the HTTP response from either URI-M or URI-TG(URI-M). It will contain an entry of the form:

```
<URI-R>;rel="original"
```

This is straightforward for clients and servers to implement, and is mandatory in the Memento specification.

Given URI-R, it is now necessary to discover the interval of time, from $\min(t_i)$ to $\max(t_i)$, in which URI-M's representation was served from URI-R. This is the first and last time that the representation was delivered from URI-R. The `Content-Datetime` response header from URI-M contains the lowest value of $t_i$ that the archival server knows, however the actual value of $\min(t_i)$ is likely to be lower for crawler based web archives.

The lower bound of $\min(t_i)$ is the timestamp for the previous Memento, URI-$M_{i-1}$. The upper bound of $\max(t_i)$ is the timestamp for the next Memento URI-$M_{i+1}$. The actual values for $\min(t_i)$ and $\max(t_i)$ will be somewhere between $t_{i-1}$ and $t_i$, and $t_i$ and $t_{i+1}$ respectively. These URI-M are listed in the `Link` header, in entries of the form:

```
<URI-M_{i-1}>;rel="prev-memento";datetime="..."
<URI-M_{i-1}>;rel="next-memento";datetime="..."
```

As an approximation, when given only this information, we could take a point half way between the current and previous Mementos as $\min(t_i)$, and between the current and next Mementos as $\max(t_i)$. The time of the first Memento would be its own lower bound, and the current time would be $\max(t_i)$ for the most recent Memento. This is frequently going to be inaccurate; the yellow shaded areas in Figure 6 show when the prediction is correct for the example.

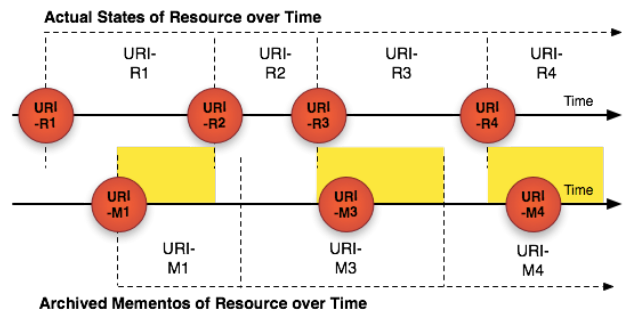

**Figure 6. Archived vs Actual State of Resources**

We now use URI-R and the time interval as search keys to discover the applicable annotations in the Annotation Collection.

For quick and dirty estimates, this approach has some merit as it can be calculated from the header information. However, better precision is required for any serious annotation implementation.

### 4.2.2 Strategies for Increasing Accuracy

The best strategy for improving the precision is to look in the TimeMap that describes the full set of Mementos for URI-R. The

TimeMap should be discoverable from a `Link` HTTP header in the Memento or TimeGate response.

URI-R can be discovered by looking in the TimeMap for a *mem:OriginalResource*, or via the SPARQL query:

```
SELECT ?uri_r WHERE { ?uri_m mem:mementoFor
?uri_r . ?uri_r a mem:OriginalResource . }
```

Transactional Archives [10] and Content Management Systems are capable of maintaining an extremely accurate knowledge of when particular representations were served. This precise information is asserted in the TimeMap using the *mem:validOver* predicate with a TimeSpan as the object, which has *mem:start* and *mem:end* properties that hold the datestamps. The SPARQL query would therefore be:

```
SELECT ?start ?end WHERE {?uri_m mem:validOver
?span . ?span mem:start ?start . ?span mem:end
?end . }
```

This will coincide almost exactly with the actual states in Figure 6, and hence the yellow shaded areas would cover practically the entire width of the diagram.

Even a crawling archive may have some additional knowledge concerning the validity period, as it may have encountered the same representation in consecutive crawls. Instead of *mem:validOver*, crawling archives use the *mem:observedOver* predicate to indicate this lack of certainty. They can also use *mem:observations* to indicate the number of observed occurrences of the representation within this period.

Further techniques to improve accuracy might include user interaction or machine learning to determine if the annotation content is likely to be applicable to the representation. We defer to the previous work on robustness of annotations in this matter.

## 5. EXPERIMENTAL EVALUATION

In order to prove the conceptual solution, a proof-of-concept system was implemented and tested.

## 5.1 Experimental Resources Used

The experimental resources used are as follows:

**Client:** The client was implemented in javascript, tailored to the Firefox browser and its support for dynamically creating SVG. The SVG objects created were used as segment descriptions and for later highlighting the appropriate region of the resource on retrieval. The implementation relied heavily on the AJAX pattern for dynamically creating, retrieving and displaying the annotations, with custom processors for SRU and RDF/XML implemented. The Djatoka [7] server was used to display images and provide zooming and panning.

It must be said that although content negotiation takes place all the time on the web, it is not possible to accomplish on demand in any of the standard web browsers. For security reasons, javascript is not allowed to add or modify request headers, making it impossible to set the `Accept-Datetime` header based on information from the annotation in this way. It is possible in a plugin, but plugin functionality is not accessible to javascript.

Until Memento gains wider acceptance and it becomes possible to directly create or manipulate headers, workarounds must be put in place to use the Memento framework from a web browser. To fill this gap, we created a server-side client that performed the interactions with the correct headers using an extended QtWebKit[9]. The URI-R and the time were given and the server-side client followed the redirections and generated a screenshot of the resulting page. The screenshot was then displayed in Firefox for the user, rendered by Djatoka.

**Content Server:** Blogger[10] was used as the Content Server where the Annotation Transcriptions were initially uploaded directly from the javascript client. It was also chosen due to being a well known, third party system with easy account creation. The transcriptions were serialized in RDFa (RDF embedded in HTML) and are hence readable by humans as well as machines.

**Annotation Database:** The Annotation Collection database was implemented using the Cheshire3 [21] system, extended to process RDF graphs using rdflib[11]. As the complete Transcription is needed by the client, SRU 1.2 was used as the search protocol rather than the graph-centric SPARQL language which would have only returned individual triples. RDF/XML was used as the default recordSchema, however the Talis[12] and GData[13] specifications for RDF in JSON were also implemented for ease of client processing.

**Target Resources:** The resources created for the original Memento experiment were reused as the Targets, comprising of automatically generated images and encapsulating HTML page, available at http://lanlsource.lanl.gov/hello. Oft-modified pages from Wikipedia were also annotated as a third party, verifiable source. Further tests involved resources in the Internet Archive.

**Content Resources:** The content resources were plain text files.

**TimeGates, Memento Servers:** The resources from the Memento experiment are archived within a transactional archive that also implements the TimeGate specification. The Wikipedia articles are maintained by their database, and exposed as the history for each page. However, as Wikipedia does not support Memento, it was necessary to use a proxy over top of the Wikipedia API to resolve the URI plus time into the appropriate version. The Internet Archive was also made Memento-compliant via a screen-scraping proxy.

## 5.2 Experimental Process and Results

After implementing the systems required and selecting appropriate resources, the experimental evaluation was conducted.

### 5.2.1 Mementos for a Given Annotation

To step through an example evaluation, on the 23rd of January the day's still empty current events page in Wikipedia was annotated with a segment and the content "No news is good news". The client posted this annotation to Blogger, from where it was collected and indexed by the Cheshire3 system. This process is depicted below at the top of Figure 7.

The following day, after many changes to the Wikipedia page had occurred, the client was used both with and without Memento enabled to reconstruct the annotation. The client generated a list of the required resources (content and target URIs), and, when Memento was enabled, their appropriate timestamps from the *oac:when* property on the Context nodes.

---

[9] http://doc.trolltech.com/4.4/qtwebkit.html

[10] http://www.blogger.com/

[11] http://www.rdflib.net/

[12] http://n2.talis.com/wiki/RDF_JSON_Specification

[13] http://code.google.com/apis/gdata/docs/json.html

When testing without making use of the the *oac:when* information, as shown in the middle section of Figure 7, the page was retrieved in its current state. There was now more than a screen-full of news items posted, which made the "No News" annotation irrelevant.

When time was taken into account, with the assistance of the server-side client, the annotation client requested a Memento using the Original Resource's URI and the timestamp from *oac:when*. The Memento was displayed by the client along with the Content, and the result, as depicted in the lower third of Figure 7, is a faithfully reconstructed Annotation.

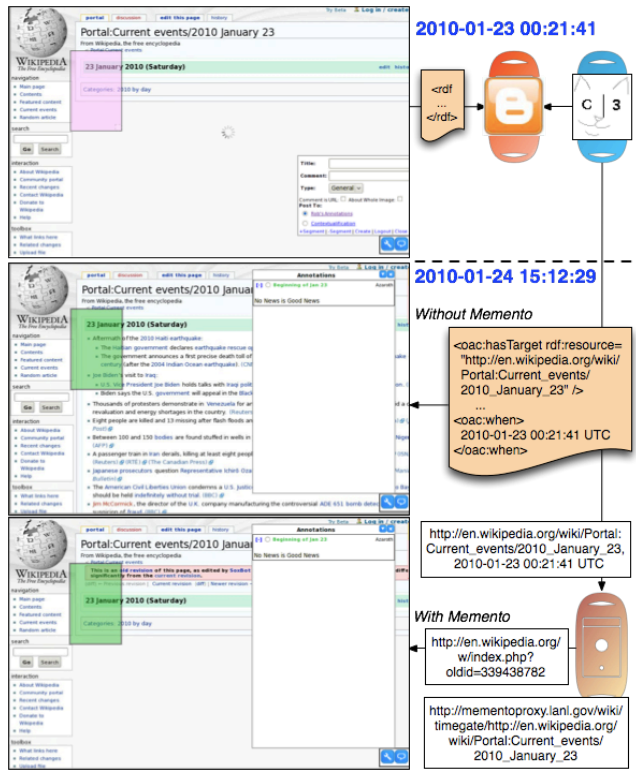

**Figure 7. Example Interactions for Question 1**

Generally, all of the resources that could be rendered without relying on resolving any other resource, such as images, worked perfectly. The same representation that was annotated was retrieved in every case. Segments of simple resources, such as overlaying SVG areas on top of images also worked perfectly in all situations, as the region of the representation was exactly where it was when the annotation was created.

All of the HTML pages and their constituent images and stylesheets, when served from either the transactional archive or from Wikipedia also worked perfectly. The Wikipedia representations were not 100% identical, as advertisements change, and a pink bar, visible in Figure 7, is inserted to say that it is an old version, however the correct content of the article was successfully retrieved and displayed for the annotation.

In further testing, a majority of the pages stored in crawler-based web archives also worked correctly, however some were mis-rendered when the archive had failed to collect all of the necessary resources, with stylesheets and javascript files being the main causes. This meant that the correct HTML content was retrieved but the layout was incorrect. Segments based on XPath or content fragments would work as expected, however coordinate based fragments did not overlay the intended part of the page.

Overall, the evaluation showed that the information provided by an OAC Annotation was sufficient to ensure that, with support for Memento, the original representation could be reconstructed.

### 5.2.2 Annotations for a Given Memento

To work through an example evaluation for the second question's solution, we first annotated the same, changing resource (http://lanlsource.lanl.gov/hello) over three days in January. These annotation creation events are shown in the top row of screenshots in Figure 8. Each annotation content describes the different t-shirt which is being worn on the particular day (white, black with super hero, black with W words, respectively).

The Transcriptions for these Annotations were stored in Blogger, and the Annotation Collection then harvested and indexed them.

On the 25th of January, given the URI for the Memento, (http://mementoarchive.lanl.gov/store/ta/201001220 10002/http://lanlsource.lanl.gov/hello) of the page as of the 22nd of January, the server-side client retrieved and inspected the Content-Datetime and Link headers to discover the URI and timestamp of the Original Resource, as well as the datetimes of the Mementos adjacent in time to the current one. The URI and time interval, accurate in this case because it came from a transactional archive, were passed to the annotation client, which then searched the Annotation Collection using its SRU interface with this information as search keys. The matching annotations were then displayed with the Memento. The description of the clothing matched the shirt worn in the image, as shown in the bottom right of Figure 8, indicating that the retrieved annotation corresponded with the Memento. The same search was done without specifying a time interval to ensure that this was not the only annotation on the resource. The bottom left of the figure shows the resulting list of annotations.

Overall, the information contained in the Memento headers returned from either a transactional archive or content management system was in all cases sufficient to retrieve all and only the annotations specific to that representation. As the content management system knew the modification times of all of the versions of the resources, and the transactional archive had the times when different representations were first served, the interval between different representations captured exactly the correct set of annotations. This information could also have been extracted from the TimeMap describing the set of Mementos available. In our experiments with crawler-based web archives, the correct annotations were also retrieved. The degree of success for this type of archive depends entirely on the relative coverage of Mementos for the resources being annotated.

These results demonstrate that the Memento approach in compliant systems can indeed be used to determine the set of appropriate annotations for the given archived version of a resource. This prevents the client from displaying annotations irrelevant to a target resource.

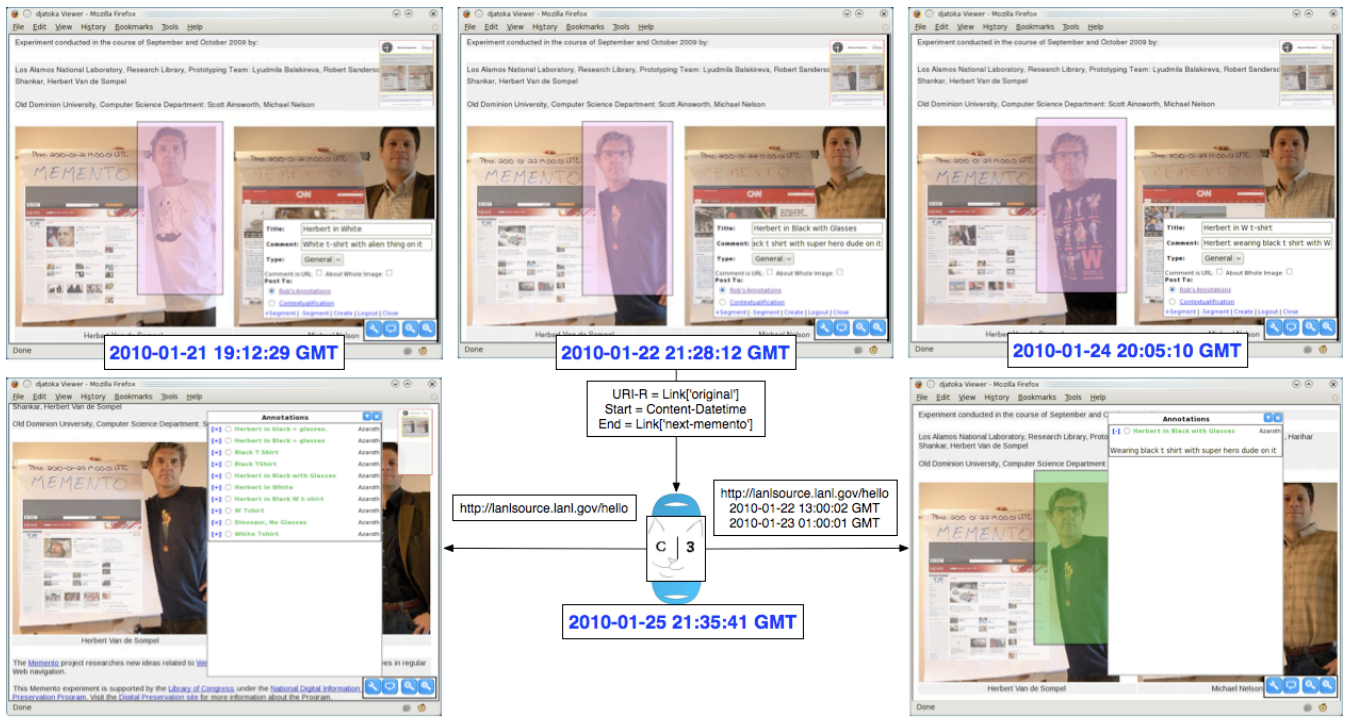

**Figure 8. Example Interactions for Question 2**

## 6. DISCUSSION

There is, as always, much valuable work to be done that this initial research does not address. The research described did not attempt to discover if the annotation was relevant across multiple representations, but not all representations. Support for this capability could be added to the OAC model with an end point in time for the Annotation's validity, however it seems unrealistic to expect a client to add this information after the annotation's creation. An extension service may be able to help in this regard.

The restrictions imposed by the browser that make it impossible to implement Memento, or any other system relying on HTTP headers, in plain javascript create a serious issue for acceptance and future work. A good solution must be found, either by engaging with the browser developers or by discovering a client-side workaround. Signed plugins that are trusted by the browser to set request headers may be a generalized solution.

If the target and content servers are not Memento compliant, the client should be prepared to use known TimeGates to attempt to find archived copies rather than simply being redirected. And finally, there are no guarantees that there will be a valid Memento for any given annotated resource. In fact, it is highly unlikely unless the original server is a content management system that maintains all of the versions of the resource. This suggests that annotation clients should be proactive in ensuring that the resources involved in annotations are archived. This could be achieved, for example, by interaction with on-demand web archives. Equally, web archives could subscribe to alerting services, such as Atom feeds, for new annotations and archive the included resources.

## 7. CONCLUSIONS

As scholarship rapidly transitions to a web-based endeavor, and the scholarly discourse becomes more tightly integrated with the broader human debate that takes place on the web, digital scholarly assets conceptually undergo a metamorphosis from digital objects held by a digital library to resources – identified by a URI - set free on the web. This conceptual shift begs rethinking services that involve scholarly assets, including annotation, in terms of the web at large rather than in terms of isolated digital libraries.

Devising a web-centric annotation framework is not without its challenges, and this paper has focused on meeting a crucial requirement for scholarly annotation: robustness of annotations over time. Due to the architectural design of the web, which allows addressing resources by means of their URI, but provides no means of individually addressing their evolving representations, achieving annotation robustness is not trivial.

By combining the temporal features built into the emerging Open Annotation model, with the capability offered by the recently introduced Memento framework that allows HTTP-navigation from the URI of a resource to archived versions thereof, this paper has proposed a conceptual solution that allows achieving a web-centric annotation framework that provides guarantees regarding annotation robustness. The conceptual solution has been experimentally explored. Under the assumption that archived versions of resources exist, the experiments were successful yet hindered by constraints on the implementation of annotation clients resulting from the inability to manipulate HTTP headers.

Hence, we can conclude that it is possible to devise a time-robust, web-centric annotation framework. As the Memento approach, and more generally the REST paradigm, becomes increasingly widespread it must be expected that the current client implementation constraints will be alleviated. More generally, we conclude that the findings of this paper suggest that many services that include scholarly assets that are currently implemented in terms of isolated digital libraries can be reframed in a web-centric perspective.

## ACKNOWLEDGMENTS
OAC was funded by the Andrew W. Mellon Foundation, which contributed to this research.  Memento was partly funded by the Library of Congress, which also contributed to this research. The authors would like to thank Scott Ainsworth, Luda Balakireva, Tim Cole, Bernhard Haslhofer, Jane Hunter, Cliff Lynch, Michael Nelson, Doug Reside and Harihar Shankar for their contributions to OAC and Memento, and directly and/or indirectly to this paper.